\documentstyle[aps,amssymb,preprint,graphics]{revtex}

\begin{document}

\title{Comparative Studies of Lensing Methods}

\author{Thomas P.  Kling, Ezra T.  Newman, and Alejandro Perez \\ University of Pittsburgh,
Pittsburgh, PA, USA}

\date{\today} \maketitle

\tightenlines

\begin{abstract}

Predictions of the standard thin lens approximation and a new iterative
approach to gravitational lensing are compared with an ``exact'' approach
in simple test cases involving one or two lenses.  We show that the thin
lens and iterative approaches are remarkably accurate in predicting time
delays, source positions and image magnifications for a single monopole
lens and combinations of two monopole lenses.   In the cases studied, the 
iterative method provided greater accuracy than the thin lens method. We 
also study the accuracy of a ``2 lens, single lens plane model,'' where two 
monopole lenses colinear with the observer are modeled by a mass 
distribution in a single lens plane lying between them. We see that this 
model can lead to large inaccuracies in physically meaningful situations. 

{\emph{A previous version of this paper was published as {\rm{Phys.~Rev.~D, 
{\bf{62}}, 024025, (2000)}} with errors in the computation of two lens 
comparisons. This paper corrects these errors and presents new conclusions 
which differ from the previous version.}} 

\end{abstract}

\section{Introduction}

In this paper, three different approaches to gravitational lensing are compared in test cases
with one or two lenses.  The three approaches include a recently introduced exact
approach~\cite{FN}, which we will take as representing correct values in our comparisons, the
thin lens approximation commonly used in lensing, and a new iterative technique whose zeroth
iterate is given by the thin lens approximation~\cite{KNP}.  An outline of these three
methods will be given in Section~II.

A recent paper by Frittelli, Kling, and Newman~\cite{FKN} discusses lensing in a
Schwarzschild geometry and compares the exact approach with the standard thin lens
approximation, a second order thin lens approximation, and a strong-field version of the thin
lens approximation introduced by Virbhadra and Ellis~\cite{VE}.  In Schwarzschild spacetimes,
it was found that the first and second order thin lens approximations fail dramatically when
the light rays encounter strong gravitational fields.  However, the strong-field version of
Virbhadra and Ellis, which is a hybrid lensing approach using some exact and some thin lens
ideas, performed remarkably well in predicting the observation angle of a given source for a
given observer and lens, even when the light ray circles the lens many times.  In addition,
it was found that the errors in the time delays predicted by the thin lens approximation
compared with the values predicted by the exact method were small for cases resembling
current observational scenarios, but, as the impact parameter was reduced, the error
introduced by using the thin lens approximation became appreciable.

Here we are interested in studying the accuracy of the thin lens and iterative approaches to
gravitational lensing in more detail.  Specifically, we wish to address two questions:

\begin{enumerate}

\item Are the intrinsic errors introduced by using the thin lens approximation of comparable
size to the observational errors present today or in the near future?

\item Are the corrections for lens structure, given in terms of higher multipole moments or
some mass distribution function, utilized by the thin lens method, comparable to the inherent
errors in the thin lens approximation?

\end{enumerate}

\noindent With each of these questions, we are also interested in seeing if the iterative
approach provides a significant improvement in accuracy.  We will not examine the accuracy of
other approximate techniques in the literature~\cite{pyne1,pyne2}.

At this point, we do not wish to imply that the work presented here should be taken as a
serious attempt at modeling real lens systems.  Our intention is to study 
the accuracy of the thin lens and iterative methods in simple test cases 
and to determine whether more detailed studies should be undertaken.

To address these issues, several test cases involving one or two lenses will be examined.
Three different lens configurations will be compared:

\begin{description}

\item[i.]  a single spherically symmetric lens,

\item[ii.]  two identical lenses spatially collinear with the observer in two different lens
planes,

\item[iii.]  two identical lenses located in a single lens plane.

\end{description}

\noindent For each of these scenarios, we will consider several comparisons:

\begin{description}

\item[$1.$] the source location predicted by each method given the same lens configuration,
observer, and observation angle,

\item[$2.$] the time delays between two rays predicted by each method given the same lens
configuration, observer, and (two) observation angles.

\item[$3.$] the magnification (relative to an unlensed ray) predicted by each method given
the same lens configuration, observer, and observation angle.

\end{description}

\noindent Section~III will be devoted to these comparisons.  The issue of the accuracy of the
thin lens approximation was raised in Chapter~9 of~\cite{ehlers}, where a footnote reference
is given to the work of P.~Haines.  One motivation of this paper is to extend the studies
cited there.

In general, we find that, in the observational regime, the thin lens and iterative approaches
are very accurate when applied to lensing by a single monopole lens and 
combination of two monopole lenses.  In most cases studied, the iterative 
method provided only a minimal improvement over the thin lens method. 
However, when two lenses in separate, but closely spaced, lens planes are 
modeled in the thin lens approximation by a mass distribution compressed into 
one lens plane, observationally significant errors were found. 

\section{Lensing Approaches} \label{approaches}

In this section, we give brief outlines of the thin lens, exact and iterative approaches to
lensing.  For convenience, we will set $G = c = 1$ in our equations, although, in our final
comparisons we will return to physical units (meters, days).  Our convention for the
signature of the metric is $(+, -, -, -)$.

\subsection{The thin lens approximation} \label{tlapproach}

The thin lens approximation is the standard approach to gravitational lensing used by
astrophysicists.  In this subsection, we will only outline its basic premises and indicate
several particular assumptions which we will use in our comparisons in Section~III.  Very
thorough and pedagogical presentations of the thin lens methodology and its applications can
be found in the excellent book by Falco, Schneider and Ehlers~\cite{ehlers} and a paper by
Blandford and Narayan~\cite{BN}.

In the standard approach, the lens is treated as a weak perturbation of a background
spacetime.  For convenience, two kinds of spatial planes are introduced:  a source plane,
where potential sources lie, and lens planes containing lensing bodies.

\begin{figure} \begin{center} \scalebox{0.8}{\includegraphics{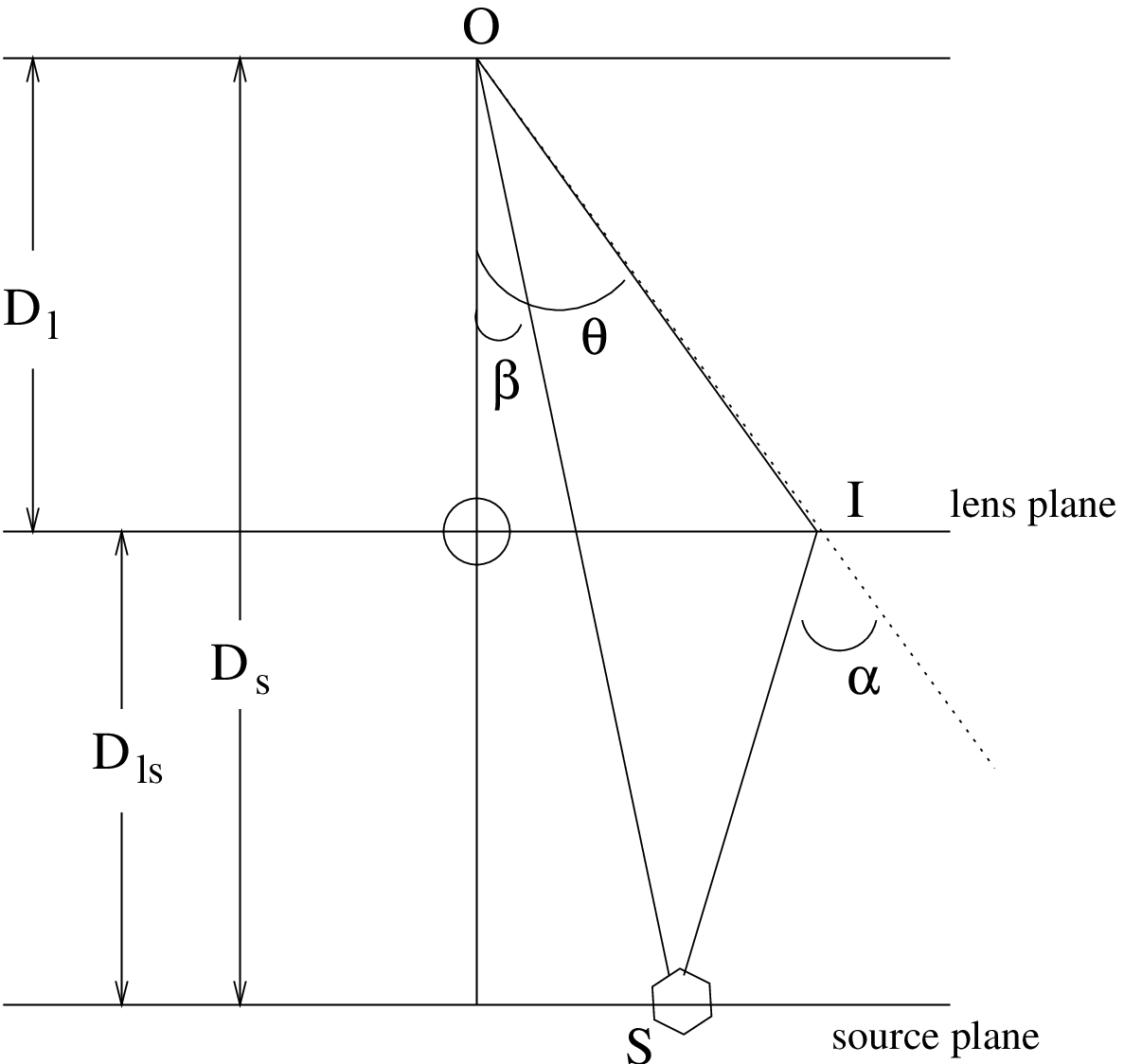}} \end{center}
\caption{Schematic illustration of the thin lens method for a single lens.}
\label{thinlensfig} \end{figure}

Kinematically possible paths from an observer at the point O to a source at a point S in the
source plane will be connected, piecewise smooth segments of geodesics in the background
space.  For example, if there is one lens plane, the trajectories from S to O will pass
through the lens plane at the image point I, as shown in Fig.~\ref{thinlensfig}.  The paths
from S to I and from I to O are geodesics in the background metric with no influence from the
lens.  The only influence of the lens on the trajectory occurs at I, where the direction of
the geodesic is instantaneously changed by an amount determined by a bending angle.  The thin
lens bending angle is a function of a mass distribution in the lens plane and the point I.

For a single, spherically symmetric lens with mass $m$, the bending angle, $\alpha$, is given
by

\begin{equation} \alpha = \frac{4\,m}{| \vec{\xi}_o |}, \label{schwarzbending} \end{equation}

\noindent where $| \vec{\xi}_o |$ is the magnitude of the two dimensional vector in the lens
plane locating the point I relative to the lens.  This bending angle is the first order
bending angle obtained in a Schwarzschild spacetime between the future and past asymptotes of
a null ray connecting points at future and past null infinity.  If there are many monopole
lenses in the plane located at $\vec{\xi}_i$ with masses $m_i$, the bending angle is given by
adding the first order contributions of the individual bending angles

\begin{equation} \vec{\alpha} = \sum_{i=1}^{n} \frac{4\,m_i}{|\vec{\xi}_i - \vec{\xi}_o|^2}
(\vec{\xi}_i - \vec{\xi}_o).  \label{multibending} \end{equation}

\noindent The vectorial bending angle, $\vec{\alpha}$, gives the two dimensional bending
angle in the lens plane.  Note that if there is more that one lens plane, the bending angle
in each plane will be influenced only by the lenses in that plane.  Continuous mass
distributions are obtained by replacing the summation in Eq.~(\ref{multibending}) by an
integral and integrating a mass distribution over the entire lens plane.

If the mass distribution is known, the entire thin lens trajectory can be codified into a
lens equation.  If $\vec{\theta}$ is an angle locating the image of a lensed source and
$\vec{\beta}$ is the ``observation angle in the absence of the lens'' given in
Fig.~\ref{thinlensfig}, the lens equation for one lens plane is

\begin{equation} \vec{\beta} = \vec{\theta} - \frac{D_{ls}}{D_s} \vec{\alpha}.
\label{tllens} \end{equation}

An important quantity in lensing is the time which elapses between the emission of the light
ray and its interception by the observer, or the time of arrival.  Although we will not use
the ideas here, the time of arrival serves as a Fermat potential from which one can derive
the lens equation, Eq.~(\ref{tllens}).  Individual arrival times are not an observable, but
the time delay between two images of the same source is an important quantity which has been
measured in several lens systems~\cite{Harvard}.  (The time delay can be used, with other
observations, to determine the Hubble constant.)

Returning to the general case with many lenses in many lens planes, the time of arrival in
the thin lens approximation is given by

\begin{equation} t = \int_{SIO} (1 + 2\,U(l)) dl, \label{tltime} \end{equation}

\noindent where $U(l)$ is the Newtonian potential of the mass distribution, and the integral
is taken along the thin lens trajectory parameterized by an Euclidean length $l$ from the
source at $S$ to through the image point at $I$ to the observer at $O$.  For a collection of
masses, the Newtonian potential is

\begin{equation} U(l) = \sum_{i=1}^{n} \frac{m_i}{|\vec{r}(l) - \vec{r}_i|}.  \label{NewtPot}
\end{equation}

\noindent In Eq.~(\ref{NewtPot}), $\vec{r}_i$ is a three dimensional vector locating lenses
relative to some origin while $\vec{r}(l)$ is the vector locating points along the thin lens
path parameterized by a Euclidean length along the trajectory.  Time delays are given by the
difference in two such times.

In this paper, we will choose Minkowski spacetime as our background.  We do not anticipate
that there would be significant differences in our results using Robertson-Walker or
on-average Robertson-Walker metrics as the background spacetime, although we have not
examined this question.

\subsection{Exact lensing}

The key difference between the exact approach to gravitational lensing and the thin lens
approximation is that, in the exact approach, the lens is fully incorporated into a metric
satisfying Einstein's equations.  In this way, no background / lens splitting is introduced
and no quantities defined in the thin lens method as ``in the absence of a lens'' have
meaning.

Lensing information in the exact approach is obtained by integrating the null geodesic
equations of the metric~\cite{FN}.  A particular parametric form of the null geodesic
equations are defined to be the lens and time of arrival equations.  It can be shown that in
a Schwarzschild spacetime, the lens and time of arrival equations can be expressed
parametrically in local, timelike coordinates as

\begin{eqnarray} t &=& T(R, \theta_o, \phi_o, x^a_o) \label{extime} \\ x^i &=& X^i(R,
\theta_o, \phi_o, x^a_o) \label{exlens} \end{eqnarray}

\noindent where $(t, x^i)$ label points on the past light cone of an observer located at
$x^a_o$~\cite{FKN}.  The two ``angular parameters,'' $(\theta_o, \phi_o)$, represent the
direction on an observer's celestial sphere where the image is observed, and $R$ can be taken
as a physical distance to the source such as the angular-diameter distance.
Equation~(\ref{extime}) is defined to be the exact time of arrival equation while
Eqs.~(\ref{exlens}) are the exact lens equations.

The first comparison which we will consider in this paper is lensing in Schwarzschild
spacetimes.  The time of arrival and lens equations for a Schwarzschild spacetime can be
found in closed form by integrating the null geodesics of the Schwarzschild metric using its
symmetries~\cite{KN}; however, in the current work, we will not use these closed form
expressions for the lens and time of arrival equations.  Instead, we will solve the null
geodesic equations by forming a Hamiltonian,

\begin{equation} H(x^c, p_c) = g^{ab}(x^c)\,p_a\,p_b, \label{exHamiltonian} \end{equation}

\noindent and numerically solving Hamilton's equations of motion.  Null geodesics are
obtained when the initial conditions, $(x^a_o, p_a^o)$, satisfy

\begin{equation} H(x^c_o, p_c^o) = 0.  \label{nullcond} \end{equation}

\noindent We will refer to the numerical integrations of the geodesic equations as the
``exact'' approach or method.

By performing a coordinate transformation,

\begin{eqnarray} t &=& t \nonumber \\ r &=& \sqrt{x^2 + y^2 + z^2}\nonumber \\ \theta &=&
\arctan{\frac{\sqrt{x^2 + y^2}}{z}} \nonumber \\ \phi &=& \arctan{\frac{y}{x}},
\label{coordtrans} \end{eqnarray}

\noindent we can write the Schwarzschild metric as

\begin{equation} g^{ab} (x^a) = \left( \begin{array}{cccc} 1/(1 - \frac{2m}{r}) & 0 & 0 & 0
\\ 0 & -1 + \frac{ 2 m x^2}{r^3} & \frac{ 2m x y}{r^3} & \frac{2m x z}{r^3} \\ 0 & \frac{ 2 m
x y}{r^3} & -1 +\frac{2m y^2}{r^3} & \frac{ 2m y z}{r^3} \\ 0 & \frac{ 2 m x z }{r^3} &
\frac{ 2m y z}{r^3} & -1 + \frac{2m z^2}{r^3} \end{array} \right).  \label{exactmetric}
\end{equation}

\noindent These coordinates are useful for comparing the ``exact'' method to the thin lens
and iterative methods.  Hence, we solve Hamilton's equations of motion in a $(t, x, y, z)$
coordinate system which is adapted to comparisons with the thin lens and iterative
approaches.

We are also interested in comparing the three lensing methodologies in situations with more
than one lens.  Since there are no exact solutions to Einstein's equations to meet these
lensing configurations, one can not apply the exact methodology.  For these cases, we will
consider approximate metrics whose null geodesics are solved using Hamilton's equations of
motion.  In these cases, we will call the numerical solution to Hamilton's equations for null
geodesics of the approximate metric the ``exact'' time of arrival and lens equations.

\subsection{Iterative approach}\label{itlensing}

The iterative approach seeks to improve upon the thin lens approximation and can be applied
to any approximate solution to Einstein's equations which is close to a spacetime in which
the exact method can be employed.  In this paper, we will focus on the first iterate only,
although higher iterates can be obtained.  Details on the iterative approach, including
equations for the first iterate method applied to a Schwarzschild spacetime, can be found
in~\cite{KNP}.

The general method is to assume that the spacetime of interest, $(M,\,g^{ab})$, is close to
some spacetime, $(M,\,g^{ab}_o)$, where the geodesic equations can be solved exactly.  This
means that we can write the metric $g^{ab}$ as

\[ g^{ab}(x^c) = g^{ab}_o(x^c) + h^{ab}(x^c) \]

\noindent where the components of $h^{ab}(x^c)$ are small.

One begins the iterative method by forming the Hamiltonians in both spacetimes,

\begin{equation} H(x^c,\, p_c) = g^{ab}(x^c)\,p_a\,p_b = g_o^{ab}(x^c)\,p_a\,p_b +
h^{ab}(x^c)\,p_a\,p_b \label{itHamiltonian} \end{equation}

\begin{equation} H_o(x^c,\,p_c) = g^{ab}_o(x^c)\,p_a\,p_b, \label{tlHamiltonian}
\end{equation}

\noindent and solving the Hamilton-Jacobi equation in $(M, \, g^{ab}_o)$:

\begin{equation} g^{ab}_o(x^c) \frac{\partial F}{\partial x^a}\frac{\partial F}{\partial x^b}
+ \frac{\partial F}{\partial \lambda} = 0.\label{Ham-Jac} \end{equation}

\noindent In spacetimes in which the geodesic equations can be solved, one can always find a
solution to the Hamilton-Jacobi equation of the form

\begin{equation} F = F_o (x^c, P_c, \lambda) \end{equation}

\noindent in which $\lambda$ is a parameter and $P_c$ are four constants.  This function can
be taken as the generating function for a parameter dependent, canonical transformation,

\begin{equation} (x^c, p_c) \rightarrow (X^c, P_c).  \label{coordtrans2} \end{equation}

If $(M,\, g^{ab}_o)$ is taken to be Minkowski spacetime, $(M, \eta^{ab})$, the solution to
the Hamilton-Jacobi equation, Eq.~(\ref{Ham-Jac}) is

\begin{equation} F_o (x^c, P_c, \lambda) = x^a P_a - \eta^{ab}P_{a}\,P_{b}\lambda,
\label{MinkF} \end{equation}

\noindent and the canonical transformation to the coordinates $(X^c, P_c)$ is given by

\begin{eqnarray} x^a &=& X^a - 2 \eta^{ab} P_b \, \lambda \label{xXtrans}\\ p_a &=& P_a.
\label{pPtrans} \end{eqnarray}

\noindent When this canonical transformation is applied to the Hamiltonian in $(M,\,g^{ab})$,
the transformed Hamiltonian takes a particularly simple form

\begin{equation} H(x^c,\, p_c) \rightarrow H'(X^c\,P_c,\,\lambda) =
h^{ab}(X^c,\,P_c,\,\lambda)P_a\,P_b.  \label{H'} \end{equation}

\noindent Hamilton's equations for geodesics in the $(X^a, P_a)$ coordinates are

\begin{eqnarray} \dot{X}^a &=& 2h^{ab}(X^c,\,P_c,\,\lambda)\,P_b + \left(\frac{\partial
h^{bc}}{\partial P_a}(X^c,\,P_c,\,\lambda)\right)\,P_b\,P_c \equiv \Xi^a (X^a, P_a, \lambda)
\nonumber \\ \dot{P}_a &=& - \left(\frac{\partial h^{bc}}{\partial
X^a}(X^c,\,P_c,\,\lambda)\right)\,P_b\,P_c \equiv \Pi_a (X^a, P_a, \lambda).  \label{hameqns}
\end{eqnarray}

\noindent No approximations have been made to obtain these equations.

We wish to solve Hamilton's equations, Eqs.~(\ref{hameqns}), by iteration.  For the zeroth
iterate, we must specify eight functions

\begin{eqnarray} X^a_0 &=& X^a_0(X^a_o, P_a^o, \lambda) \nonumber \\ P_a^0 &=& P_a^0(X^a_o,
P_a^o, \lambda), \end{eqnarray}

\noindent and substitute these functions of $\lambda$ and initial conditions, $x^a_o = X^a_o$
and $p_a^o = P_a^o$, into the right hand side of Hamilton's equations, Eqs.~(\ref{hameqns}).
(Care should be taken to choose the eight functions serving as the zeroth iterate close to
the true description of the path of the null geodesic.)  The first iterate is obtained by
direct integration on $\lambda$:

\begin{eqnarray} X^a_1 (X^a_o, P_a^o, \lambda) &=& X^a_o + \int_0^\lambda \, d\lambda' \,
\Xi^a (X^a_0, P_a^0, \lambda') \nonumber \\ P_a^1 (X^a_o, P_a^o, \lambda) &=& P_a^o +
\int_0^\lambda \, d\lambda' \, \Pi_a (X^a_0, P_a^0, \lambda').  \label{ithameqns}
\end{eqnarray}

\noindent Likewise, the $n$th iterate is obtained by placing $(X_{n-1}^a (X^a_o, P_a^o,
\lambda), P_a^{n-1} (X^a_o, P_a^o, \lambda))$ into the right hand side of
Eqs.~(\ref{hameqns}) and integrating up.

The $n$th iterate solution to Hamilton's equations in the original spacetime coordinates,
$(x^a, p_a)$, is obtained by placing $(X_n^a (X^a_o, P_a^o, \lambda), P_a^n (X^a_o, P_a^o,
\lambda))$ into the canonical transformation, Eq.~(\ref{xXtrans})~and~(\ref{pPtrans}):

\begin{eqnarray} x_n^a (X^a_o, P_a^o, \lambda) &=&X^a_n (X^a_o, P_a^o, \lambda) -
2\eta^{ab}\,P_b^n (X^a_o, P_a^o, \lambda)\,\lambda \label{nthx} \\ p_a (X^a_o, P_a^o,
\lambda) &=& P_a^n (X^a_o, P_a^o, \lambda).  \end{eqnarray}

\noindent In these equations, $(X^a_o = x^a_o, P_a^o = p_a^o)$ are the initial values for the
approximate geodesic.  When these initial conditions satisfy the null condition on the
Hamiltonian in Eq.~(\ref{itHamiltonian}),

\begin{equation} H(X^a_o = x^a_o, P_a^o = p_a^o, \lambda = 0) = g^{00}(x^a_o)\,(p^o_0)^2 +
g^{ij}(x^a_o) \,p_i^o\,p_j^o = 0, \label{nullit} \end{equation}

\noindent the geodesics are approximately null.  Solving Eq.~(\ref{nullit}) for $p_0^o$
yields

\begin{equation} p_0^o = \sqrt{\frac{-g^{ij}(x^i_o) \,p_i^o\,p_j^o}{g^{00}(x^a_o)}}.
\label{setp0}\end{equation}

\noindent We will make use of this expression below.

To make a deeper connection with lensing, we will take the thin lens path as the zeroth
iterate.  As an example, we show the explicit zeroth iterate for the case of one spherical
lens in one lens plane.

For one spherically symmetric lens, the thin lens path would be given by

\begin{eqnarray} x_{(1)}^i &=& x^i_o - 2 \delta^{ij} \, p_j^o \, \lambda \quad\quad
0<\lambda<\lambda_1 \nonumber \\ x_{(2)}^i &=& x_{(1max)}^i - 2 \delta^{ij} \,
{\tilde{p}}^{o}_j \, \lambda \quad\quad 0<\lambda, \label{tlpath} \end{eqnarray}

\noindent where $x_{(1)}^i$ describes the first leg (to the lens plane), $x_{(2)}^i$
describes the second leg away from the lens plane and $(x^i_o, p_i^o)$ are constant initial
conditions.  If we locate the observer at $-|z_o|$ on the $-\hat z$ axis, we can use
spherical symmetry to consider geodesics in the $\hat x$-$\hat z$ plane.  Then the $z = 0$
plane is the lens plane, and the value of $\lambda_1$ is

\begin{equation} \lambda_1 = \frac{ |z_o|}{2\,p_z^o}.  \end{equation}

\noindent We also have that

\begin{equation} x_{(1max)}^i = x^i_o - \frac{\delta^{ij} \, p_j^o \, |z_o|}{p_z^o}.
\end{equation}

Using the bending angle, $\alpha = 2 m / |x_{(1max)}^i|$, the ${\tilde{p}}^{o}_j$ will be
determined up to scaling through the Minkowski spatial inner product between $p_j^o$ and
${\tilde{p}}_j^o$:

\begin{equation} \cos{\alpha} = \frac{(p_x^o)({\tilde{p}}_x^o) +
(p_z^o)({\tilde{p}}_z^o)}{|p_i^o|\,|{\tilde{p}}^{o}_i|}.  \label{alphaeqn} \end{equation}

\noindent If we multiply and divide the right hand side in Eq.~(\ref{alphaeqn}) by $1/(p_z^o
\, {\tilde{p}}_z^o)$ and define

\[ p = \frac{p_x^o} {p_z^o} \quad\quad {\rm {and}} \quad\quad{\tilde{p}} =
\frac{{\tilde{p}}_x^o }{{\tilde{p}}_z^o}, \]

\noindent Eq.~(\ref{alphaeqn}) is equivalent to a quadratic equation for $\tilde p$ in terms
of $p$ and $\alpha$.  Solving this equation gives

\begin{equation} {\tilde p} = \frac{p \pm \cos{\alpha} \sin{\alpha}\sqrt{(1 +
p^2)^2}}{\cos^2{\alpha} - p^2 \sin^2{\alpha}}.  \label{tildep} \end{equation}

We now return to the issue of the value of $p_0^o = P_0^o$.  In this paper, the iterative
method will only be applied to stationary spacetimes.  Hence, the timelike coordinate, $t$,
will not appear in the original Hamiltonian, Eq.~(\ref{itHamiltonian}), and $T$, the
canonically transformed variable, will not appear in the transformed Hamiltonian,
Eq.~(\ref{H'}).  As $T$ is cyclic, the time equation, $({\dot T})$, will separate from the
spatial equations.  Moreover, because $T$ does not appear in the Hamiltonian,

\[ \dot{P}_0 = - \frac{\partial H'}{\partial T} \]

\noindent will be identically zero so that $P_0 = p_0$ is constant.  The value of this
constant must be chosen to make the trajectory null, as in Eq.~(\ref{setp0}).  However, there
is an inherent scaling freedom of the null vector which permits us to define new
four-momenta, $p_a^{'o}$ which are a constant multiple of the old one.

It is customary to fix the freedom in the scaling of the null vector by {\it{defining}}
$p_0^{o'} = 1$.  Then in the case of a single monopole lens, the equation

\begin{equation} p_0^o \equiv 1 = \sqrt{\frac{-g^{ij}(x^a_o) \,p_i^o\,p_i^o}{g^{00}(x^a_o)}},
\label{setpz}\end{equation}

\noindent gives an equation for the initial $p_z^o$ in terms of the $p_x^o$ and the initial
point.

As we take the thin lens trajectory as the zeroth iterate, we will choose to set the value of
$p_0^o$ to one at the initial point (the observer, $O$) and again in each lens plane.  In
this way, $p_0^o = {\tilde{p}}_0^o = 1$, and the relative scaling of $(p_x^o, p_z^o)$ and
$({\tilde{p}}_x^o, {\tilde{p}}_z^o)$ is uniquely determined using Eq.~(\ref{setpz}).

Summarizing, the spatial part of the zeroth iterate for a one lens system is given by the
thin lens path, Eq.~(\ref{tlpath}).  The observer is located at the initial point $x^a_o$.
For a given value of $p_x^o$, $p_z^o$ is uniquely determined by the condition $p_0^o = 1$,
from Eq.~(\ref{setpz}).  The new $({\tilde{p}}_x^o, {\tilde{p}}_z^o)$ are determined using
the bending angle as in Eq.~(\ref{tildep}) and the condition $\tilde{p}_0^o =1$.

So far, our discussion has only considered cases with axial symmetry and one lens plane.  If
there is more than one lens plane, the procedure we have described above is extended to each
lens plane.  If there is no axial symmetry, there will be a complementary equation to the
bending angle relation, Eq.~(\ref{alphaeqn}), which can be used to fix the three spatial
components of the momenta in an analogous way to what we have presented here.

In the cases we will study, the time coordinate is cyclic and the time equation in the
original phase space variables is simply the integral equation

\begin{equation} t = t_o + \int 2\,p_0^o (1 + 2 \, U(\,x^a(\lambda)\,)) d\lambda,
\label{ittime}\end{equation}

\noindent where $U(r)$ is the Newtonian potential of the mass distribution.  The integral is
taken over the path as a function of the parameter $\lambda$.  The first iterate time is
obtained when the path inserted into Eq.~(\ref{ittime}) is the zeroth order, thin lens path,
$x^a(\lambda) = x^a_0(\lambda)$.  Hence the first iterate time is precisely the value
obtained in the standard thin lens approximation.  However, as we will find the first iterate
trajectories, we can also find the second iterate time values by evaluating the integrand in
Eq.~(\ref{ittime}) along the first iterate, $x^a(\lambda) = x^a_1(\lambda)$.

A conceptual problem arises in computing the iterative time of arrival if one uses the
formula

\begin{equation} t_n = t_o + 2\int\, p_0~d\lambda~~ (\,1\, +\, U(\,x^a_{n-1}(\lambda)\,)\,)
\label{tn1} \end{equation}

\noindent for the $n$th iterate time.  In general, the parameter $\lambda$ should be an
affine parameter along a null geodesic.  We note that if Hamilton's equations are solved
exactly, fixing $p_0^o$ as in Eq.~(\ref{setp0}) ensures that the value of the Hamiltonian
will be zero at all points along the path and $\lambda$ will be an affine parameter.
However, in the iterative method, the geodesic equations are not solved exactly, and the
value of the Hamiltonian will slowly drift away from zero.  Hence, as $\lambda$ grows, it
fails to be an affine parameter along a null geodesic.

A way to force the Hamiltonian to be zero, and hence $\lambda$ to be null affine parameter,
is to allow $p_0$ to be a function along the trajectory given by

\begin{equation} p_0 = p_0^n (\lambda) = \sqrt{\frac{- g^{ij} (x^a_n)\, p_i^n\,
p_j^n}{g^{00}(x^a_n)}}, \label{varyp0} \end{equation}

\noindent where $(x^a_n, p_a^n)$ are the $n$th iterate values.  This proposal leads to a
conflict between maintaining the null value of the Hamiltonian along the $n$th iterate
trajectory

\begin{equation} H(x^a_n, p_a^n, \lambda) = 0 \label{prob1} \end{equation}

\noindent and Hamilton's equation,

\begin{equation} {\dot{p}}_0 = 0.\label{prob2} \end{equation}

Since we are dealing with static cases, a consistent proposed solution is to take $p_0$ as
constant in the spatial Hamilton's equations, but allow $p_0$ to vary as in
Eq.~(\ref{varyp0}) when computing times.  This solution disentangles the two competing
problems given in Eq.~(\ref{prob1}) and Eq.~(\ref{prob2}) by obeying Hamilton's equations
when integrating the spatial part of the geodesic and also obeying the null condition in
computing the times (which is very sensitive to integrating over an affine parameter).

Hence, when comparing the iterative time delays to the ``exact'' and thin lens delays, we
will consider the time of arrivals given by

\begin{equation} t_2 = t_o + \int \, 2\, p_0^1 (\lambda) (1 + 2 \, U(r_1)) \, d\lambda.
\label{finittime} \end{equation}

\noindent We will show that this formula gives very accurate predictions for time delays in
our comparisons.

\section{Comparison of Lensing Approaches} \label{compsec}

In this section we present the results obtained in the comparison of time delays, source
locations and image magnifications predicted by the thin lens approximation and the iterative
method for several different lens models.  The comparisons are made with respect to the
numerical integration of the exact geodesic equations of the spacetime metric defined by the
given model.  For the iterative method, we will be interested in the first iterate only.

There are four subsections in this section.  In the first, we give details regarding the
comparisons we will be discussing.  We then group our comparisons in three sets.  First, we
consider lensing by a single spherical lens, or the Schwarzschild geometry.  Next, we
consider multiple lensing by single monopole lenses collinear with the observer in different
lens planes.  Finally, we consider lensing by two lenses in the same lens plane.  In our
plots, all angles are given in arc seconds and all times are given in days.

\subsection{Notes about comparisons}

We will be comparing the predictions of the thin lens, iterative and ``exact'' approaches to
lensing for time delays, source locations and image magnification.  In this subsection, we
give some details about how these comparisons are performed.  We will refer to the spatial
axis connecting the lens and observer as the optical axis.

First, we note that the exact solution to the geodesic equation produces an infinite number
of images~\cite{FKN}, but that the thin lens approximation predicts only two of these images
for the case of a single lens (referred to as primary images).  This feature is shared by the
first iterate method, since it corresponds to the next step in the perturbative series whose
zeroth order is given by thin lens approximation (the first iterate method should give
sensible predictions as long as its trajectory remains close to the thin lens trajectory).
However, it is not difficult to choose the two primary exact images corresponding to the thin
lens and first iterate images because these primary images are widely separated (in angular
location) from the secondary images which circle the lens one time.

Throughout our comparisons, we will refer to an observation angle, $\theta$.  This angle is
computed by taking the inner product between the spatial part of the initial momentum vector,
$p_i^o$, at the observer and the spatial vector pointing to the lens from the observer's
location, $x^a_o$.  Formally, this must be done using the spatial metric describing the
model.  However, as we will always be taking this inner product at a large distance from the
lens, it is appropriate to take the observation angle, $\theta$, as the ratio of the $\hat x$
and $\hat z$ components of the initial momentum:

\[ \theta = \frac{p_x^o}{p_z^o}.  \]

To compare the time delays predicted by the three methods, we must compute the two
trajectories from each method, integrate the arrival time function along each trajectory, and
subtract the two values we obtain.  We begin by choosing two initial angles, $(\theta_1,
\theta_2)$, one on each side of the lens.  These angles are chosen such that trajectories
with these initial conditions intersect at a reasonable distance beyond the last lens;
usually, this distance is chosen as approximately the same as the distance between the
observer and the first lens.  For each path, we compute the time of arrival and subtract the
two times to get a time delay.  We will then hold one angle, $\theta_2$, fixed while varying
$\theta_1$.  This allows us to consider the time delay as a function of $\theta_1$ for a fixed
$\theta_2$.

In practice, finding the time delays is a very difficult calculation, as the arrival times
for each trajectory will agree in roughly their first 12 digits (at our scales).  The
comparison between the methods is even more difficult, as the time delays from the three
methods tend to agree to about four digits.  Hence, to resolve a difference between the thin
lens, iterative, and ``exact'' predictions for the time delays, we must know the arrival time
to approximately 16 digits of accuracy.

To compute the source location, $\beta$, for a given observation angle, $\theta$, we choose a
value of $\theta$ and a final distance along the optical axis from the observer, $D_s$.  If
the optical axis is the $\hat z$ axis and the observer is located at $z = -|z_0|$, we then
place a plane at $z = D_s - |z_0|$ which will be the ``source plane.'' We then compute, for a
given initial condition, $\theta$, the interception point in the source plane, $\vec{\eta}$.
The value of $\beta$ is {\it{defined}} to be

\begin{equation} \beta = \frac{|\vec{\eta}|}{D_s}.  \end{equation}

\noindent We will use this definition for the source location in all three models.  As
$\theta$ is varied, we will obtain $\beta(\theta)$.

Magnifications are defined as $\partial \theta \over \partial \beta$, or the inverse slope of
the $\beta$ versus $\theta$ graph.  We will compute the magnifications from the data we
obtain in our $\beta$-$\theta$ comparisons.  Note that this magnification is not directly
observable; we discuss it only because it plays a role in the literature.  With some
additional work, we could compute, for two given images, $(\theta_1, \theta_2)$, the relative
magnification,

\[ \mu_{12} = \frac{\left(\partial \theta \over \partial \beta\right)_{\theta_2}}{\left(
\partial \theta \over\partial \beta \right)_{\theta_1}}, \]

\noindent which is observable.  We plan to return to this possible comparison in future work.

\begin{figure} \begin{center} \scalebox{.8}{\includegraphics{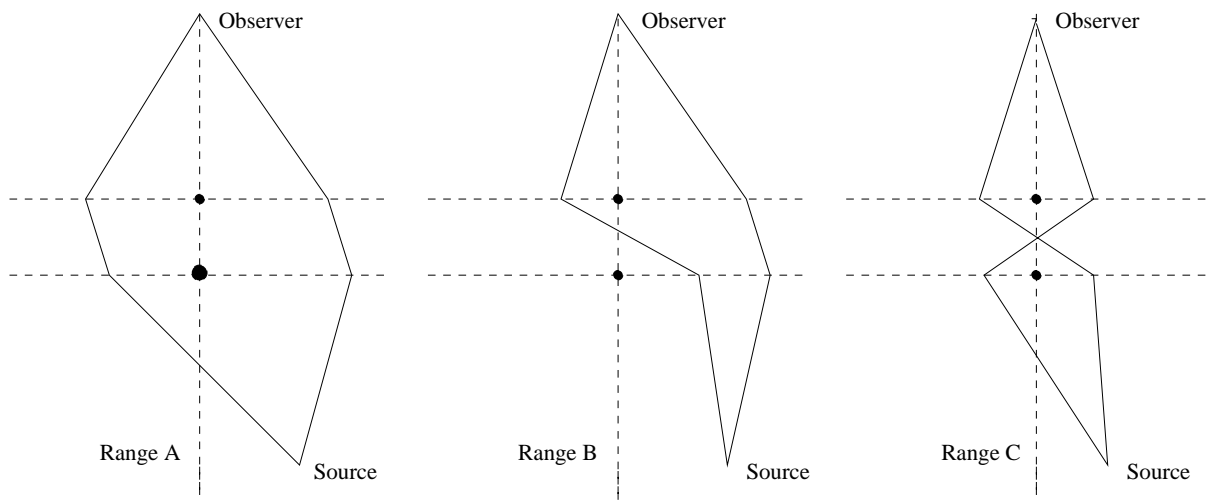}} \end{center}
\caption{Schematic illustration of the different lensing scenarios for the two lens model.}
\label{2thin} \end{figure}

In the two lens models, the thin lens approximation predicts four images.  Therefore, in the
calculation of the time delays we can distinguish three qualitatively different situations.
As it is illustrated in Fig.~\ref{2thin}, there is a range for $\theta_1$ and $\theta_2$ for
which the two rays do not cross the optical axis between the two lenses before converging at
the observer's position (range A), a range for which only one of the rays crosses the axis
between the lenses (range B), and, finally, a range in which both of the rays cross the axis
before they meet at the observer's position (range C).  When looking at magnifications and
image positions, we can compare the three methods in two cases:  1) rays which do not cross
between the two lenses and 2) rays which do cross the optical axis between the lenses.

\subsection{One spherical lens}

Here we discuss the comparisons between the first iterate, thin lens and ``exact''
predictions for various observables when there is one spherically symmetric lens.  For the
``exact'' predictions, we consider the numerical integration of the geodesic equations of the
exact Schwarzschild metric, as specified in Section~II.B.  As mentioned, Minkowski spacetime
will be considered the background spacetime for the iterative and thin lens approaches.

For our comparisons, we will take a lens $4$ billion light years (ly) away from the observer
with a mass of approximately $2.5 \times 10^{12}~M_\odot$.  While we are not concerned with 
cosmological models here, this distance scale is reasonable for current lensing studies. 

\begin{figure} \begin{center} \scalebox{.72}{\includegraphics{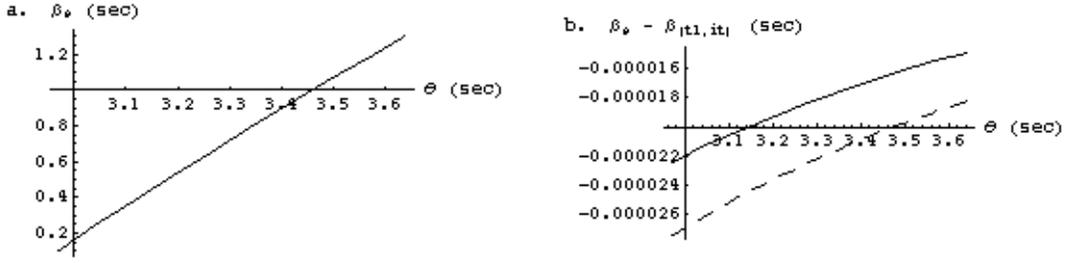}} \end{center}
\caption{a.~Exact angular position of the source, $\beta$, as a function of the image
position, $\theta$, in arc sec.  b.~Error in the angular position of the source as a function
of $\theta$.  The first iterate error ($\beta_{ex}-\beta_{it}$) and the thin lens
approximation error ($\beta_{ex}-\beta_{tl}$) are represented by a smooth line and a dashed
line respectively, in arc sec.}  \label{beta1} \end{figure}

\begin{figure} \begin{center} \scalebox{.72}{\includegraphics{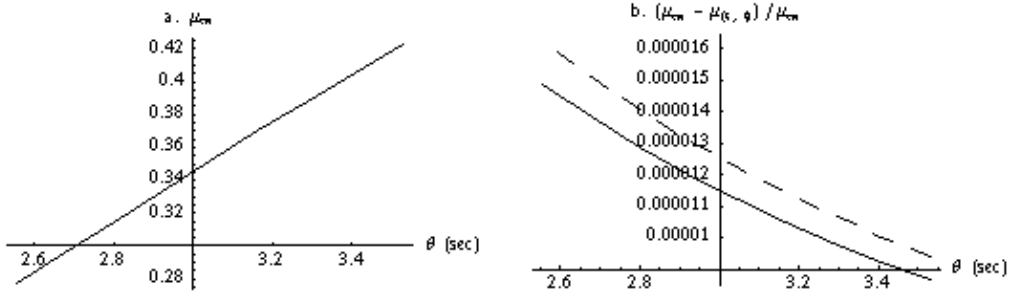}} \end{center}
\caption{a.~Exact magnification as a function of $\theta$.  b.~Relative errors in the
magnifications predicted by the first iterate and thin lens are represented by smooth and
dashed lines, respectively, as a function of $\theta$.}  \label{mag1} \end{figure}

In Fig.~\ref{beta1}a we show the position of the source, $\beta$, as a function of the image
position, $\theta$, (see Fig.~\ref{thinlensfig}) calculated using the exact numerical
integration of the geodesics equations.  The range in $\theta$ has been chosen in agreement
with observed image angles in systems with similar characteristics as the one represented by
our model, and the value of $D_s$ has been set equal to the observer-lens spacing.  The
absolute error in $\beta$ between the exact numerical integration and the thin lens
approximation and the first iterate method are shown in Fig.~\ref{beta1}b.  The discrepancies
between the two methods are of about $10^{-5}$ arc sec.  In Fig~\ref{mag1}a, we show the
``exact'' magnification $\mu = {\partial \theta \over \partial \beta}$ as a function of the
image position $\theta$.  The relative error,

\[ \Delta \mu_{tl, it} = \frac{\mu_{ex} - \mu_{tl, it}}{\mu_{ex}}, \]

\noindent in the predicted magnification by the two approximate methods is shown in
Fig~\ref{mag1}b.  The errors here are very small.

In the case of a single spherically symmetric lens, we were not able to resolve the
difference in the time delay error between the thin lens and first iterate trajectories.  Our
calculations showed that this error was indeed quite small, and that for a lens with mass 
$2.5
\times 10^{12} M_\odot$ at $4$ billion ly from the observer, the error in the thin
lens and iterative methods was less than $0.2$ days when the ``exact'' time delay was $400$
days.

\subsection{Two lenses in different lens planes}

In this subsection, we will consider lensing by two identical lenses.  We choose to 
study two different cases: when the distance between the two lenses is on 
the same order of magnitude as the distance between the observer and the 
first lens and when the distance between the two lenses is small compared 
to the distance between the observer and the first lens. We will examine the case where 
the lenses are far apart first.

\begin{center} {\bf{$1)$ Large separation}}\end{center}

\begin{figure} \begin{center} \scalebox{.72}{\includegraphics{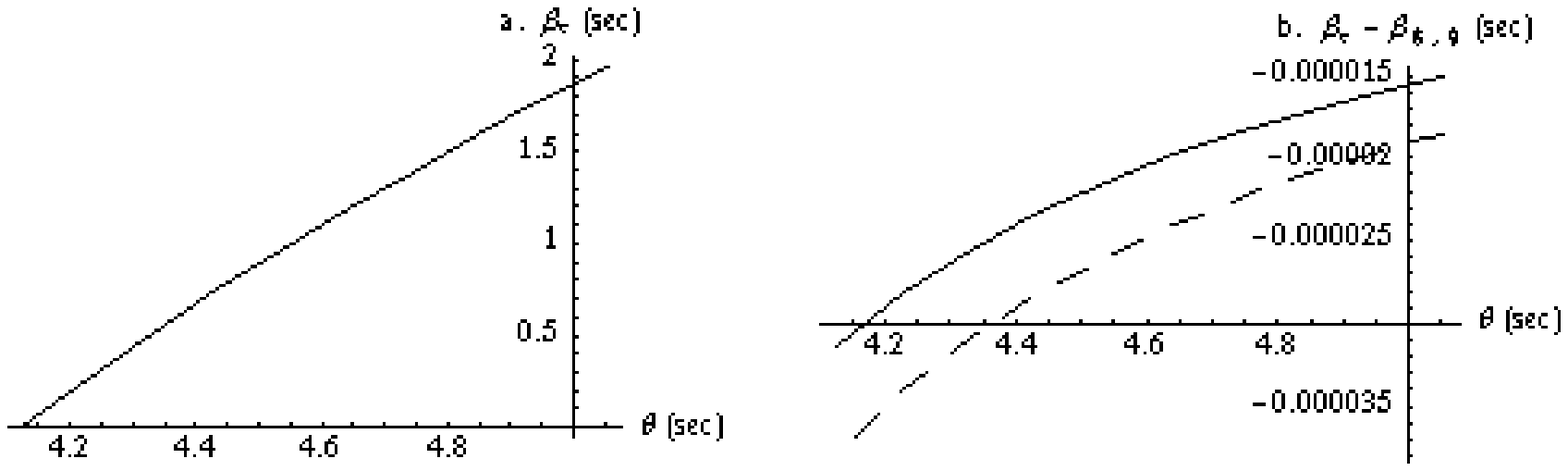}} \end{center}
\caption{a.~Exact angular position of the source, $\beta$, as a function of the image
position for two lenses separated by a distance equal to the separation of the first lens and
observer when the light ray does not cross the optical axis.  b.~Errors in the angular
position of the source predicted by the first iterate and thin lens, represented by smooth
and dashed lines, respectively.}  \label{beta2oo} \end{figure}

\begin{figure} \begin{center} \scalebox{.72}{\includegraphics{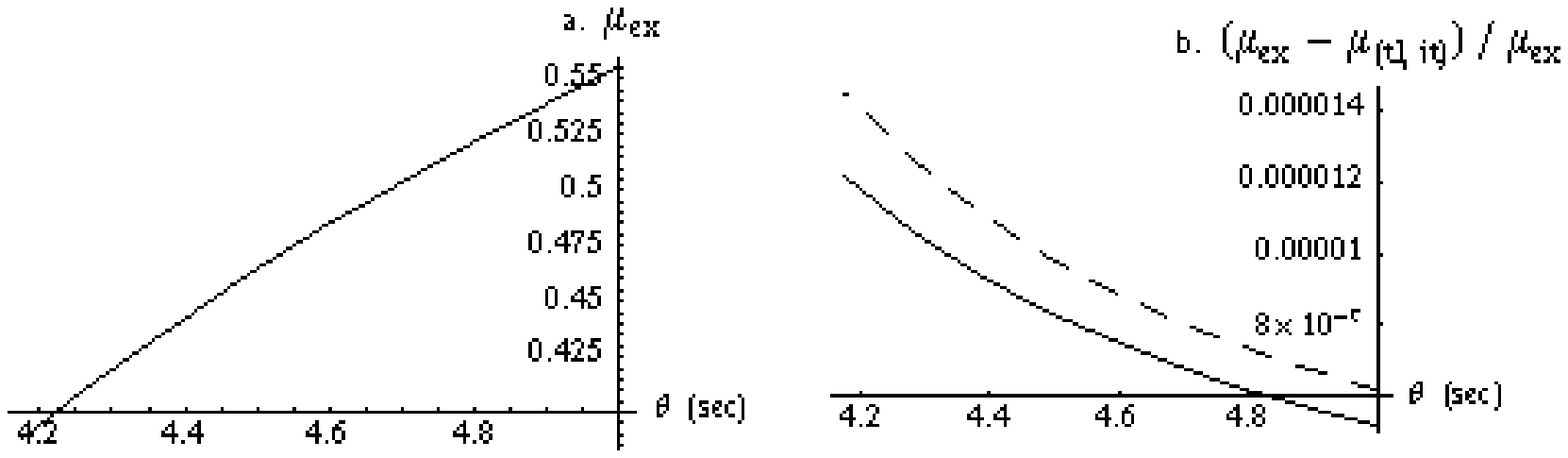}} \end{center}
\caption{a.  Exact magnification as a function of $\theta$ for two lenses separated by a
distance equal to the separation of the first lens and observer when the light ray does not
cross the optical axis.  b.~Relative errors in the magnifications predicted by the first
iterate and thin lens, represented by smooth and dashed lines, respectively.}
\label{mags2oo} \end{figure}

In our first comparisons, we consider an observer $3$ billion ly away from the first
lens and set the distance between the two lenses equal to the distance between the observer
and the first lens.  The mass of the two lenses is approximately $1.9 
\times 10^{12} M_\odot$. Figure~\ref{beta2oo}a shows the ``exact'' image location, $\beta$, as a
function of observation angle, $\theta$, when the light ray does not cross between two lenses
and $D_s = 9$ billion ly, or three times the spacing between the first lens and
observer.  As before, the error in the thin lens and first iterate methods, shown in
Fig.~\ref{beta2oo}b, are small.  At a fairly large observation angle, $4.3''$ from the
optical axis, the error in the thin lens method is about $3 \times 10^{-5}$ arc sec.  The 
``exact'' magnification and errors in the thin lens and iterative methods for this lens 
configuration are shown in Fig.~\ref{mags2oo}.

\begin{figure} \begin{center} \scalebox{.72}{\includegraphics{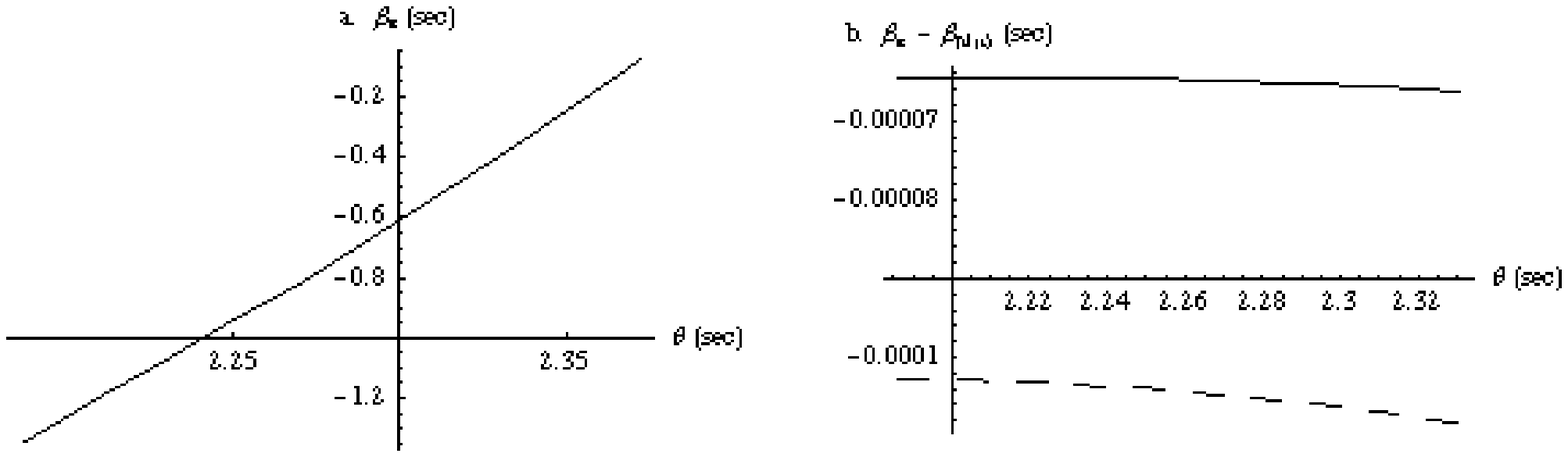}} \end{center}
\caption{a.~Exact angular position of the source, $\beta$, as a function of the image
position for two lenses separated by a distance equal to the separation of the first lens and
observer when the light ray crosses the optical axis.  b.~Errors in the angular position of
the source predicted by the first iterate and thin lens, represented by smooth and dashed
lines, respectively.}  \label{beta2zz} \end{figure}

\begin{figure} \begin{center} \scalebox{.72}{\includegraphics{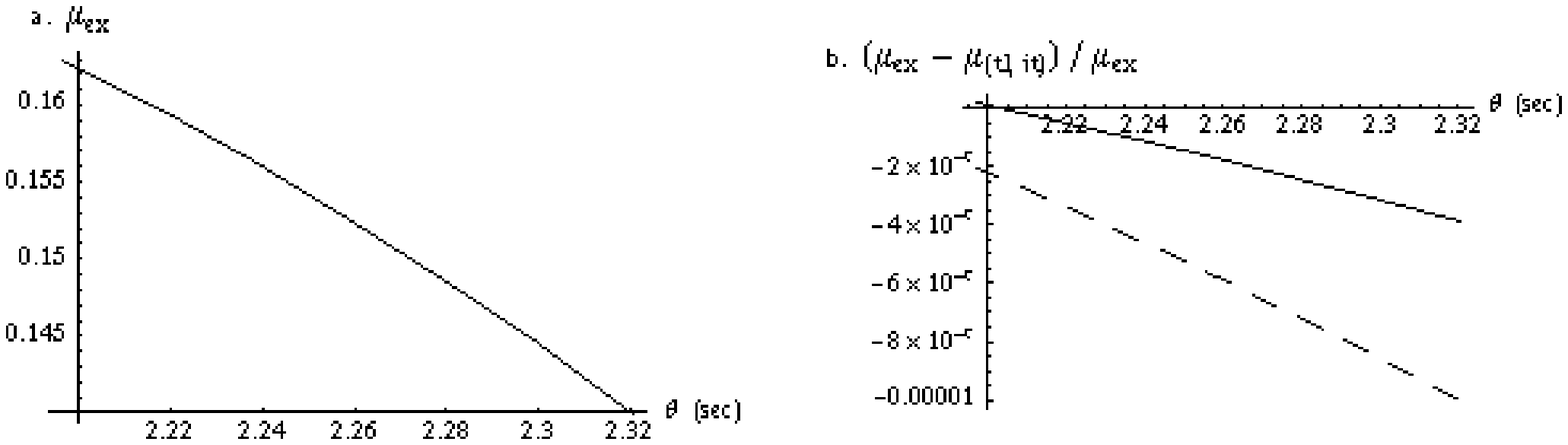}} \end{center}
\caption{a.~Exact magnification as a function of $\theta$ for two lenses separated by a
distance equal to the separation of the first lens and observer when the light ray crosses
the optical axis.  b.~Relative errors in the magnifications predicted by the first iterate
and thin lens, represented by smooth and dashed lines, respectively.}  \label{mag2zz}
\end{figure}

For the same lensing configuration, the ``exact'' source location and the error in the thin
lens and iterative methods when the light ray crosses the optical axis is shown in
Fig.~\ref{beta2zz}.  In this case, the light ray passes much closer to the lens, and we see
that the first iterate is slightly better than the thin lens in predicting the source
location.  The corresponding magnifications are plotted in Fig.~\ref{mag2zz}.

With the lens separation equal to the distance between the observer and 
the first lens, the
time delays predicted by the thin lens and iterative method are very accurate.  As in the one
lens case, we were unable to resolve a difference in the time delays due to the high
precision required in any of the three possible ray combinations from Fig.~\ref{2thin}.  Our
calculations show that the error in the thin lens and first iterate methods was less than
$0.1$ days for a time delay around $500$ days.

\begin{center} {\bf{$2)$ Small separation}}\end{center}

Similar results were obtained when the distance between the two lenses was small.
As an example, we will examine the case where the observer and two lenses lie along the same
optical axis, the mass of each lens is $2.5 \times 10^{12} M_\odot$, the distance to the first
lens from the observer is $4$ billion ly and the distance between the two
lenses is $4$ million ly.  This second distance is about twice the distance
between our galaxy and Andromeda, so that our lensing configuration represents a pair of
lenses at roughly the same distance from the observer.  Thus, we may think of this
example as corresponding to direct lensing by two members of the same group.  In this case,
it makes sense to choose $D_s$ as twice the distance to the first lens, $D_s = 8$ billion ly.

When the lenses are so close together, it does not make sense to consider rays crossing
between the two lenses; to pass between the lenses, the observation angle must be less than
$0.125''$.  Because this observation angle does not look reasonable, we 
will not compare the thin lens, iterative and exact methods in this paper 
in this range of $\theta$, 
although we note that in an analogous case of microlensing such comparisons may be 
important.

\begin{figure} \begin{center} \scalebox{.72}{\includegraphics{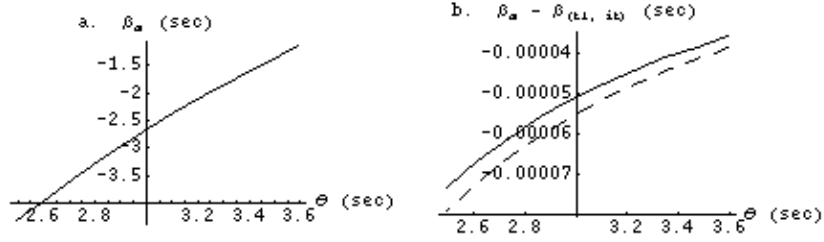}} \end{center}
\caption{a.~Exact angular position of the source, $\beta$, as a function of the image
position for two lenses separated by a small distance compared with the separation of the
first lens and observer.  b.~Errors in the angular position of the source predicted by the
first iterate and thin lens, represented by smooth and dashed lines, respectively.}
\label{beta2oo7} \end{figure}

As in our previous comparisons,  we show the ``exact'' source angle as a
function of the observation angle and the error in the thin lens and first iterate
approaches in Fig.~\ref{beta2oo7} for the case where the two lenses are 
close together.  We note that the error in the thin lens method is 
approximately $\beta_{ex} - \beta_{tl} \approx 5.5 \times 10^{-5}$ arc sec, while the error in 
the iterative  method is approximately $\beta_{ex} - \beta_{it} \approx 5.0 
\times 10^{-5}$ arc sec.
These errors are somewhat larger than the errors when the lens planes are 
widely separated. 

\begin{figure} \begin{center} \scalebox{.72}{\includegraphics{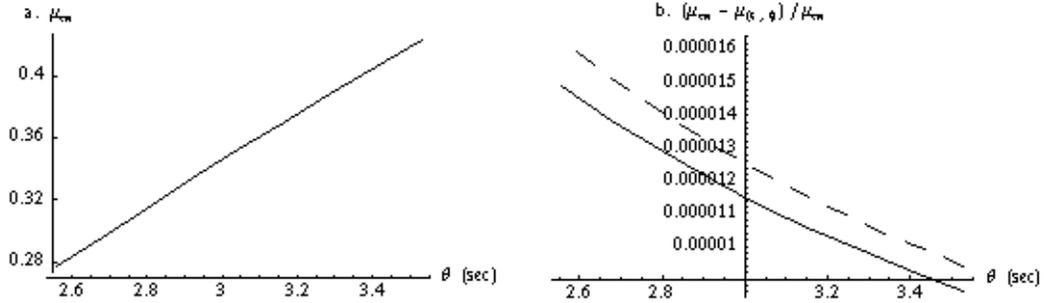}} \end{center}
\caption{a.~Exact magnification as a function of $\theta$ for two lenses separated by a small
distance compared with the separation of the first lens and observer.  b.~Relative errors in
the magnifications predicted by the first iterate and thin lens, represented by smooth and
dashed lines, respectively.}  \label{mags2oo7} \end{figure}

A similar result is found in the magnifications, shown in Fig.~\ref{mags2oo7}.  Here, we note
that the inaccuracy in both methods is about twice the inaccuracy in the 
case where the lenses are widely separated.

We did not detect an observable error in the time delays for the thin lens 
or iterative methods for this scenario. For a ``exact'' time delay of $400$ 
days, the thin lens and iterative methods were accurate to less than $0.1$ 
days for observation angles around $2.475''$.

\begin{figure} \begin{center} \scalebox{.72}{\includegraphics{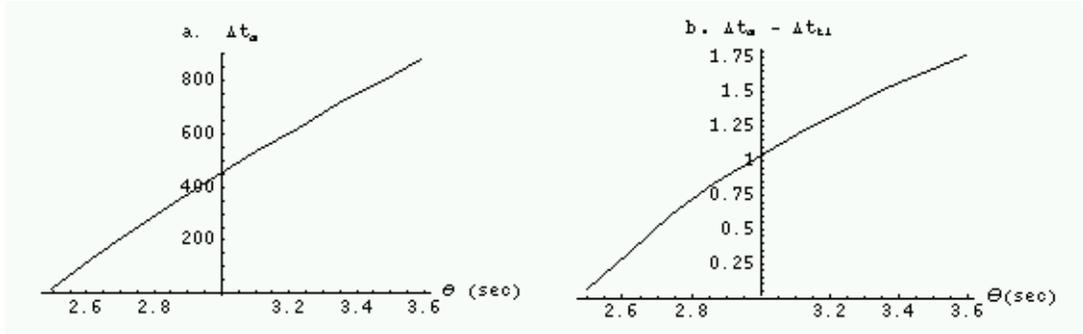}} \end{center}
\caption{a.~Exact time delay as a function of $\theta_1$ for two lenses separated by a small
distance compared with the separation of the first lens and observer when $\theta_2 =
2.475''$.  b.~Error in the time delay predicted by the 
thin lens approximation when the two lenses are compressed into one lens 
plane.}  \label{time2oo7} \end{figure}

Because the lenses are so close together, one may think that it is appropriate to treat them
as a single lens located in the middle of the two with a total mass equal to the sum of the
values of the two masses. We will refer to this model as the ``2 lens, single lens 
plane model.'' When the thin lens approximation is applied to the ``2 lens, single 
lens plane model,'' the error in the time delay is significant at this separation.  The
curve in Fig.~\ref{time2oo7}b represents the error in the time delay predicted by the thin 
lens approximation when the lensing configuration is treated as one lens with the total 
mass located directly between the two lenses. Here, one observation angle 
was fixed at $2.475''$ and the second angle varies as shown. The exact time 
delay is plotted in Fig.~\ref{time2oo7}a.

The errors present in the ``2 lens, single lens plane model'' are 
significant and are of the same order of magnitude as current 
observational abilities. Hence, it appears that lens structure extending 
along the optical axis connecting the observer and first lens can 
adversely affect the accuracy of the thin lens methodology at today's 
observational level when the structure is collapsed into a single lens 
plane. We will discuss this issue further in 
Sec.~\ref{discussion}.

\subsection{Two lenses in the same lens plane}

As a final comparison, we consider two lenses in the same lens plane.  Here we will take each
lens to have a mass of $1.25 \times 10^{12} M_\odot$ and will set the lens plane $4$ billion 
ly away from the observer.  This case resembles the one lens system in that the
distances are the same but the mass has been split.  We choose a separation of $4000$ ly 
between the lenses.  For our comparisons of $\beta$ and magnification, we will
take $D_s$ to be twice the spacing between the lens and observer.

\begin{figure} \begin{center} \scalebox{.72}{\includegraphics{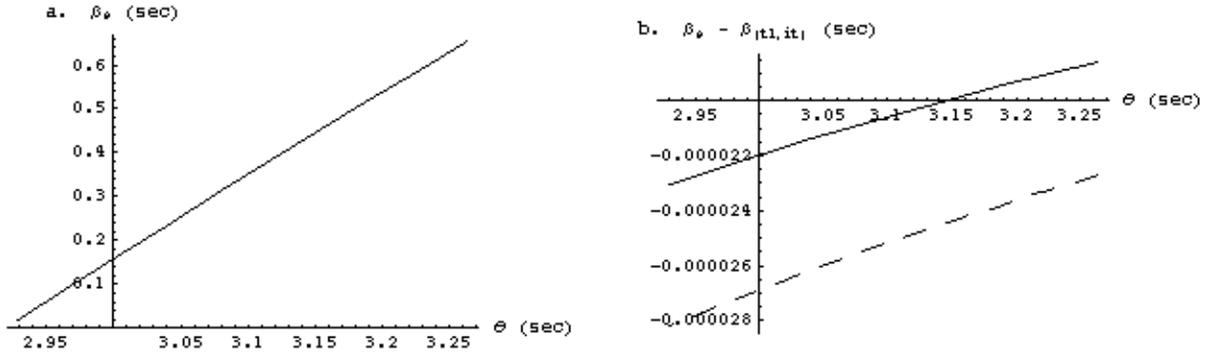}} \end{center}
\caption{a.~Exact angular position of the source $\beta$ as a function of the image position
for two lenses in the same lens plane.  b.~Errors in the angular position of the source
predicted by the first iterate and thin lens, represented by smooth and dashed lines,
respectively.}  \label{beta2lp1} \end{figure}

\begin{figure} \begin{center} \scalebox{.72}{\includegraphics{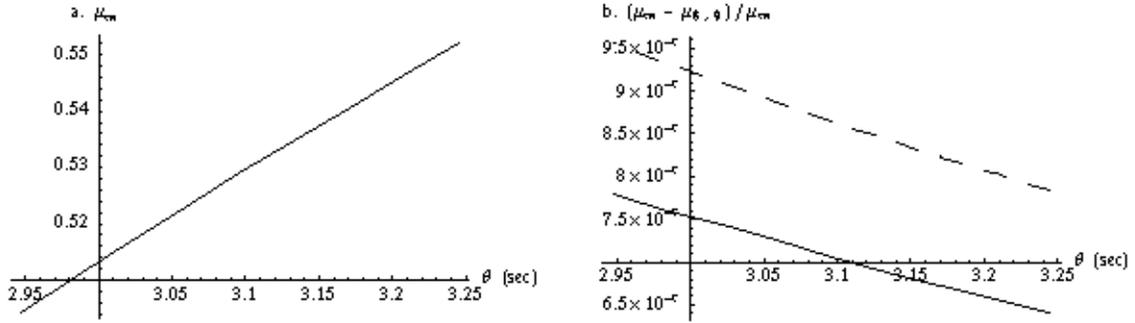}} \end{center}
\caption{a.~Exact magnification as a function of $\theta$ for two lenses in the same lens
plane.  b.~Relative errors in the magnifications predicted by the first iterate and thin
lens, represented by smooth and dashed lines, respectively.}  \label{mag2lp1} \end{figure}

Figure~\ref{beta2lp1} shows the ``exact'' plot of source angle, $\beta$, versus observation
angle, $\theta$, and the error in $\beta$ for the thin lens and iterative method.  The
``exact'' magnifications and relative errors are shown in Fig.~\ref{mag2lp1}.  As in the case
of the two lenses close together, there is a slight difference in the accuracy of the
thin lens and iterative methods.  Again, no measurable error was found in 
the time delays predicted by the thin lens and iterative methods.

\section{Discussion} \label{discussion}

We have performed a careful examination of the accuracy of two lensing approximations, the
thin lens and iterative methods, in scenarios with one and two monopole lenses.  In general,
we find that source locations, time delays and magnifications computed using the first
iterate are more accurate than those computed with the thin lens.

Both methods are accurate beyond the current level of observational error when the deflector
was a single spherical lens or two lenses in all cases tested.  The cases we
studied in this paper tended to involve rather massive lenses.  Since the thin lens and
iterative methods generally become more accurate as the mass is decreased holding the
distances the same, we feel that it is likely that lensing by objects  with 
smaller masses than those considered here will be well described by both the thin lens
and iterative methods.

It was found that when two closely separated lenses on the optical axis were modeled in the 
thin lens approximation by a mass distribution in one lens plane (the ``2 lens, single 
lens plane model''), observationally 
significant errors arise. Because these errors approach zero in the limit 
that the distance between the two lens planes goes to zero, the important 
observational question is whether there are observational scenarios where the 
depth of the mass distribution along the line of sight is too large to be 
modeled by one lens plane. 

The failure of the ``2 lens, single lens plane model'' to accurately predict observable 
quantities for some separations raises two interesting questions for 
lensing.  First, are there lensing scenarios where multiple members
of a group lens a source?  In this case, careful observational 
work needs to be done to determine the relative spacing of the members of 
the group, for if the spacing is too large, significant errors may be 
introduced.  

It seems that when there is a three dimensional mass distribution (a lens
with structure), collapsing the distribution into a single lens plane may
lead to an approximation which fails at today's level of observational
accuracy.  Further studies are needed to determine
if two dimensional continuous mass distributions of the type used in lens
modeling are affected at the same level as the collection of 
monopole lenses studied here. This issue is important because two dimensional continuous mass 
distributions are used to predict various cosmological parameters, and the 
inability to correctly model time delays, source positions and image 
magnifications will lead to inaccuracy in the prediction of fundamental 
constants from lensing.

As a second question, we are interested in how our results apply to microlensing by binary
systems.  It is estimated that nearly ten percent of microlensing events will be microlensing
by binary systems.  At various points in time, the rotating bodies will resemble either two
lenses in different lens planes or two lenses in the same lens plane, which are similar to
our case studies.  One key difference is that, in general, the mass to distance ratio in
microlensing will be smaller than the ratios we have studied.  This will tend to reduce the
error we detected in the thin lens method.  On the other hand, as the source moves across the
sky, the light rays from the source come very close to the binary lens.  We found a rather
large error in the thin lens method when the distance of closest approach was on the same
order of magnitude as the separation between the lenses.  Hence, we do not know what the
accuracy of the thin lens method will be when applied to microlensing by binaries.  We will
study this issue in future work.

In summary, we have shown that the intrinsic errors of the thin lens
approximation {\em{fail}} to approach today's level of observational error
(approximately one milli arc sec for angles in the visible band and one day
for time delays) by approximately two orders of magnitude for one or two
monopole lenses. The inherent errors in the iterative method were 
consistently smaller than the errors of the thin lens method, although 
these errors were of the same order of magnitude in almost all cases.

On the other hand, when a single lens plane is 
used to model two closely separated lenses in different lens planes, 
significant errors did arise in time delays. This
suggests that the ``2 lens, single lens plane model'' should be applied very carefully 
in observational cases; one should be careful to check that lens structure 
does not extend a significant fraction of the distance along the line of 
sight between the lens and observer. 

Even though the inherent errors in the thin lens approximation are not 
 a significant fraction of the observational errors in the cases we 
 studied, it is not inconceivable that the observational accuracy will 
 improve over time to a point where more sophisticated approaches are 
 required for modeling lens trajectories.  The iterative method provides one such improvement 
over the thin lens and seems to be accurate in all the cases we have studied.

\begin{center} {\bf{Acknowledgments}} \end{center}

\noindent The authors would like to thank David Turnshek, Al Janis, Simonetta Frittelli and
Jurgen Ehlers for their helpful advice and suggestions.  Alejandro Perez would like to thank
FUNDACION~YPF.  This work was supported under grants Phy~97-22049 and Phy~92-05109.

\eject

\end{document}